\documentclass[aps,notitlepage,twocolumn]{revtex4-1}

\usepackage{graphicx}

\usepackage{amsmath}

\usepackage{epstopdf}

\usepackage{amsfonts,amssymb}

\usepackage[utf8]{inputenc}

\usepackage{pstricks}

\usepackage{color}

\newcommand{\be}{\begin{equation}}

\newcommand{\ee}{\end{equation}}

\newcommand{\ben}{\begin{eqnarray}}

\newcommand{\een}{\end{eqnarray}}

\newcommand{\pslash}{\not{\hbox{\kern-2.3pt $p$}}}

\newcommand{\pdslash}{\not{\hbox{\kern-2pt $\partial$}}}

\begin{document}

\title{$Z_b(10610)$  in a hadronic medium}

\author{L. M. Abreu$^{a}$, F. S. Navarra$^{b,c}$, M. Nielsen$^{b,d}$ 
and A. L. Vasconcellos$^{a}$}

\affiliation{$^{a}$Instituto de F\'{i}sica, Universidade Federal da Bahia, Campus Ondina, 40170-115, Salvador, Bahia, Brazil}

\affiliation{$^{b}$Instituto de  F\'{i}sica, Universidade de S\~ao Paulo, Rua do Mat\~ao 1371, 05508-090 São Paulo, SP, Brazil}

\affiliation{$^{c}$Institut de Physique Th\'eorique, Universit\'e Paris Saclay, CEA, CNRS, F-91191, Gif-sur-Yvette, France}

\affiliation{$^{d}$SLAC Nacional Acelerator Laboratory, Stanford University, Stanford, California 94309, USA}

\begin{abstract}

In this work we investigate the $Z_b (10610)$ (called simply $Z_b$) 
abundance in the hot hadron gas produced in the late stage of heavy 
ion collisions. We use effective Lagrangians to calculate the thermally  
averaged cross sections of  $Z_b $ production in processes  such as      
$B^{(*)} + \bar{B}^{(*)} \to \pi + Z_b (10610)$    and also of its  
absorption in the corresponding inverse processes.  We then solve the 
rate equation to follow the time 
evolution of the  $Z_b$ multiplicity, and the results remarkably 
suggest that this quantity is not significantly affected by the considered 
reactions.  The number of $Z_b$'s produced at the end of the 
quark-gluon plasma phase remains constant during the hadron phase.

\end{abstract}

\maketitle




\section{Introduction}

\label{Introduction}




During the last decade the existence of exotic hadron states has been 
established \cite{revex,PDG,Chen}. In particular,  in the bottomonium 
spectrum, the charged states 
$Z_b^\pm(10610)$ (from now on called simply $Z_b$) have been observed by the Belle Collaboration  in the      
invariant mass spectra of the $\pi ^{\pm} \Upsilon (n S)\;(n=1, 2, 3) $ 
and  $\pi ^{\pm} h_b (m P)\;(m = 1, 2) $ pairs that are produced in 
association with a single charged pion in $\Upsilon (5 S)$              
decays~\cite{Bondar}. Their masses and decay widths have been estimated  to be  $m_{Z_b  ^{\pm }} = 10607.2 \pm 2.0$ MeV and   
$\Gamma _{Z_b ^{\pm } } = 18.4 \pm 2.4$ MeV, respectively~\cite{PDG}, and the favored quantum numbers are $I^G(J^P) = 1^+(1^+)$~\cite{PDG}.      
The Belle Collaboration has also found evidence of the charge neutral 
partner $Z_b (10610)$ in the Dalitz plot analysis                       
of $\Upsilon (5 S) \rightarrow \Upsilon (2 S)  \pi ^{0}  \pi ^{0}$, 
with the mass being $m_{Z_b ^{0}} = 10609 \pm 6$ MeV~\cite{Adachi:2012im}. 
Since the 
$Z_b$ is an isotriplet state,  it cannot be pure $b \bar{b}$ state, 
needing at least four quarks as minimal constituents.

The structure of these isospin triplets is still matter of debate. Due to 
the proximity to $B\bar{B}^*$ thresholds, a natural interpretation is to 
suppose that they are bound states of bottomed 
mesons~\cite{Bondar2,UFBa1,UFBaUSP1, ClevenEPJ,Voloshin,Nieves2,Zhang,Sun,Yang,Ohkoda1,Ohkoda2,Li1,Li2,ClevenPRD,Ohkoda3,Dias,Kang,Huo}. Another plausible interpretation is that they could be  compact tetraquark states, resulting from the binding of a diquark and an antidiquark~\cite{Chen}.

The determination of the structure of the $Z_b$ states (meson molecule, 
tetraquark or a mixture, including bottomonium components) requires more experimental information. The study of the better known $X(3872)$ state has led to the conclusion that the production in  hadronic colliders may be  more sensitive to the quark configuration. It has been claimed 
\cite{han,roma1,roma2} that data on  $X(3872)$ production in proton-proton collisions are incompatible with a molecular interpretation (for a different point of view, see however \cite{brat1,brat2}).

Hadronic production of exotic states can be investigated in proton - proton collisions and in nucleus-nucleus reactions both in ultraperipheral
\cite{bruno}  and central collisions \cite{exhic1,exhic2}.  Indeed, heavy ion 
collisions (HIC) allow for a higher production rate of heavy quarks. Moreover, the formation of a quark gluon plasma (QGP)  phase allows heavy quarks to move freely and recombine to form exotic states. As described in 
\cite{exhic1,exhic2}, heavy quarks coalesce to form bound states (and possibily exotic bound states) at the end of the QGP phase. After being produced, the multiquark states interact with other hadrons during the expansion and cooling of the hadronic matter. They can be destroyed in collisions with the comovinglight mesons, but they can also be produced through the inverse processes~
\cite{ChoLee1,XProd1,XProd2,XHMET}.  The final multiplicity depends on the interaction cross sections, which, in turn, depend on the spatial configuration of the quarks. Therefore, the measurement  of the $Z_b$   
multiplicity would be very useful to  determine the structure of these  
states. Theoretical studies reported in Ref.~\cite{XProd2}, for instance, 
suggest that  the $X(3872)$ multiplicity at the end 
of the QGP phase is reduced by a factor four due to the interactions with the
hadron gas. Moreover the results found in \cite{XProd2} suggest that if 
the $X(3872)$ was observed in HICs,  it would be most likely a molecular 
state.

Extending the study of the $X(3872)$ abundance mentioned above, in this work  we investigate the $Z_b$ abundance  in the hot hadron gas produced in the late stage of heavy ion collisions. We will follow the previous works on the subject, where the interactions of the $X(3872)$ with light mesons were addressed~\cite{XProd1}. As in Refs. 
\cite{ClevenEPJ,Ohkoda3,Huo,UFBa1,UFBaUSP1}, we assume that the $Z_b$ couples to $\bar{B} B^{\ast }$ and to  $ B \bar{B}$.  
We use effective Lagrangians to calculate the thermally  averaged   cross 
sections of the $Z_b$ production in processes such as   
$B^{(*)} + \bar{B}^{(*)} \to \pi + Z_b$,   and also of the $Z_b$ absorption 
in the corresponding inverse processes. We then solve the rate equation to follow the time evolution of the  $Z_b$ multiplicity  and determine how it is affected by the considered reactions.

The paper is organized as follows. In Section~\ref{CrossSection} we describe the formalism and calculate the cross sections of  $Z_b$  
production and absorption. With these results,  in Section~\ref{ThermalAvCrossSection} we calculate the thermally  averaged 
cross sections of the processes above mentioned. Then, in  Section~\ref{abundance}  we solve the rate equation and follow the time 
evolution of the  $Z_b $ abundance. Finally, in  
Section~\ref{Conclusions} we present some  concluding remarks.




\section{Cross Sections}

\label{CrossSection}



\begin{figure}[!ht]
    \centering
        \includegraphics[width=8.0cm]{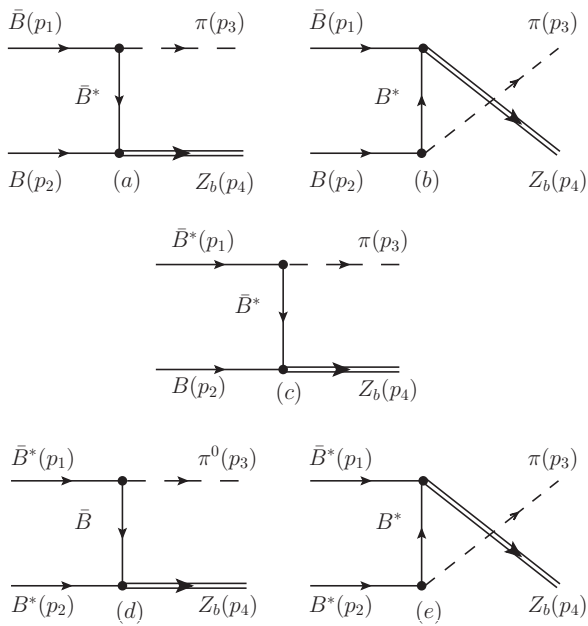}
        \caption{Diagrams contributing to the process $ \bar{B} B \rightarrow \pi Z_b $ [diagrams (a) and (b)], $ \bar{B}^* B \rightarrow \pi Z_b $ [diagram (c)] and $ \bar{B}^* B^* \rightarrow \pi Z_b $ [diagrams (d) and (e)], without specification of the charges of the particles (from Ref.~\cite{UFBaUSP1}).}
\label{DIAG1}
\end{figure}

In this section we calculate the production and absorption cross sections  
involving the $Z_b $ and pions. We focus on the processes 
$\bar{B}B \to \pi Z_b$, $\bar{B}^* B \to \pi Z_b$ and 
$\bar{B}^* B^* \to \pi Z_b$ and  the inverse reactions.  
In Fig.~\ref{DIAG1} we show the different diagrams contributing
to each process, without the specification of the particle charges.    
The cross sections for these reactions were obtained in Ref.~\cite{UFBaUSP1}  considering only the charged partner $Z_b ^{+}$. Thus, for  completeness, here we compute the total cross section including all the components of the  isospin triplet. Due to their large
multiplicity (there are about ten pions for every particle of any other 
species),  we expect that the reactions involving  
pions provide the main contributions to the study  of the $Z_b$ abundance in hot hadronic matter.


The amplitudes of the processes shown in Fig.~\ref{DIAG1} are calculated  
with the help of  effective Lagrangians based on the extended hidden $SU(4)$ local 
symmetry. For more details about explicit expressions of amplitudes and 
calculations, we refer the reader to 
Refs.~\cite{UFBaUSP1,XProd1,XProd2,Bando,Bando1,Meissner,Harada,OsetD80.114013}. The isospin-spin-averaged cross section for the processes 
$\bar{B}B, \bar{B}^*B, \bar{B}^*B^* \to \pi Z_b$
in the center of mass (CM) frame  is given by
\begin{eqnarray}
\sigma_r(s) = \frac{1}{64 \pi^2 s }  \frac{|\vec{p}_{f}|}{|\vec{p}_i|}  
\int d \Omega \overline{\sum_{S, I}}|\mathcal{M}_r(s,\theta)|^2, 
\label{eq:CrossSection}
\end{eqnarray}
where $r = 1,2,3$ label processes $\bar{B}B, \bar{B}^*B, \bar{B}^*B^*$ 
respectivelly; $\sqrt{s}$ is the CM energy;  $|\vec{p}_{i}|$ and 
$|\vec{p}_{f}|$ denote the tri-momenta of initial and final particles in  
the CM frame, respectively; the symbol $\overline{\sum_{S,I}}$ represents 
the sum over the spins and isospins of the particles in the initial and 
final state, weighted by the isospin and spin degeneracy 
factors of the two particles forming the initial state for the reaction 
$r$, i.e. \cite{XProd1,UFBaUSP1} 
\begin{eqnarray}
\overline{\sum_{S,I}}|\mathcal{M}_r|^2 & \to & 
\frac{1}{(2I_{1i,r}+1)(2I_{2i,r}+1)} \nonumber \\
&& \times \frac{1}{(2S_{1i,r}+1)(2S_{2i,r}+1)} 
\sum_{S,I}|\mathcal{M}_r|^2, \nonumber \\
\label{eq:DegeneracyFactors}
\end{eqnarray}
where 
\begin{eqnarray}
\sum_{S,I} |\mathcal{M}_r|^2 = \sum_{Q_{1i}, Q_{2i}} 
\left[\sum_{S}\left|\mathcal{M}^{(Q_{1i},Q_{2i})}\right|^2\right].
\label{eq:Sum}
\end{eqnarray}
Notice that the charges of the two particles forming the initial      
state for the processes in Fig.~\ref{DIAG1} can be combined, giving a  
total charge $Q_r = Q_{1i} + Q_{2i} = -1, 0, +1$. We have then four 
possibilities: $(0,0)$, $(-,+)$, $(-,0)$ and $(0,+)$, giving
\begin{eqnarray}
\sum_{S,I} |\mathcal{M}_r|^2 &  = & \sum_{S}\left(|\mathcal{M}^{(0,0)}_r|^2 + |\mathcal{M}^{(-,+)}_r|^2 \right. \nonumber \\                              
& & \left.  + |\mathcal{M}^{(-,0)}_r|^2 + |\mathcal{M}^{(0,+)}_r|^2\right). 
\nonumber \\
\label{eq:SumSpin}
\end{eqnarray}
As mentioned above, the expressions of the amplitudes and values of the  
coupling constants are given in Ref.~\cite{UFBaUSP1}. The uncertainties in  the couplings $g_{ZBB^*}$ will be taken into account and  the results 
discussed below will be represented by shaded regions in the plots.  
Also, in the computations of the present work we have employed  the  
isospin-averaged masses reported in Ref.~\cite{PDG}: $m_{\pi} = 137.3$  
MeV; $m_{B} = 5279.4$ MeV;  $m_{B^{*}} = 5324.8$ MeV; 
$m_{Z_b ^\pm} = 10607.2$ MeV; $m_{Z_b ^0} = 10609$ MeV.

\begin{figure}[!ht]
    \centering
       \includegraphics[{width=8.0cm}]{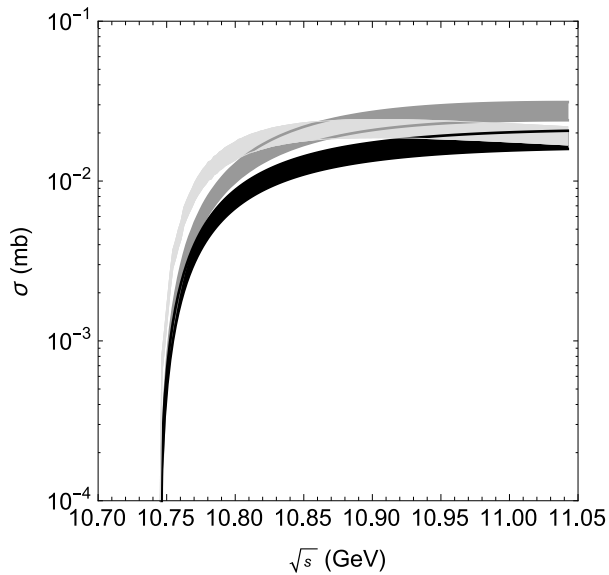} 
        \caption{
				Cross sections for the production processes $\bar{B}  B  \rightarrow  \pi Z_b $ (medium shaded region), $\bar{B}^* B \rightarrow  \pi Z_b $ (dark shaded region) and  $\bar{B}^*  B^* \rightarrow  \pi Z_b $ (light shaded region), as function of CM energy $\sqrt{s}$.
}
    \label{CrSecZb}
\end{figure}

In Fig.~\ref{CrSecZb} the total $Z_b $ production cross sections  are 
plotted as a function of the CM energy $\sqrt{s}$. 
We see that the cross sections are                  
$ \sim 6 \times 10^{-3} - 3 \times  10^{-2} $mb for                  
$ 10.80 \leq \sqrt{s} \leq 11.05 $ GeV. Also, it can be noticed that  
the three processes have cross sections with the same order of magnitude.  
Another point worthy of mention is that, due to the inclusion of the   
contributions coming from processes with all the three components of the 
isospin triplet $(Z_b ^0, Z_b^+, Z_b^- )$, we obtain cross sections that  
are bigger (by a factor about $2-3$) than those found in 
Ref.~\cite{UFBaUSP1} with only the $Z_b^+$ component.



\begin{figure}[!ht]

    \centering
            \includegraphics[{width=8.0cm}]{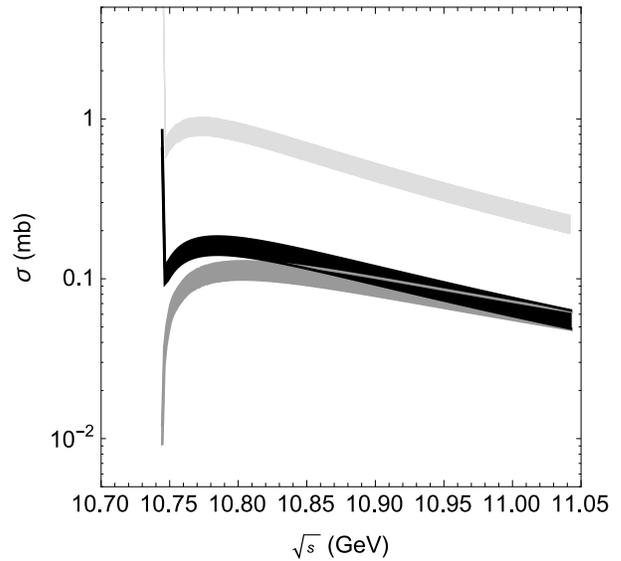} 
        \caption{
				Cross sections for the absorption processes $   \pi Z_b \rightarrow \bar{B}  B $ (medium shaded region), $  \pi Z_b \rightarrow \bar{B}^* B $ (dark shaded region) and  $  \pi Z_b \rightarrow \bar{B}^*  B^*$ (light shaded region), as function of CM energy $\sqrt{s}$.
}
    \label{CrSecZbInv}
\end{figure}

We also evaluate the cross sections related to the inverse processes.  
In Fig.~\ref{CrSecZbInv} the total $Z_b $ absorption cross sections  
are plotted as a function of the CM energy $\sqrt{s}$. 
They are found to be $\sim 5 \times 10^{-2} - 1 $ mb for                  
$ 10.80 \leq \sqrt{s} \leq 11.05 $ GeV.   As it can be seen, the reaction 
with the  final $\bar{B}^*  B^*  $ state has the largest cross section 
(by a factor about 3-10, depending on the channel) in comparison with the 
other reactions. Again, the inclusion of the contributions coming from the 
processes with all the three components $(Z_b ^0, Z_b^+, Z_b^- )$ yields 
cross sections that are bigger (by a factor about $2-3$) than those found in  
in Ref.~\cite{UFBaUSP1} with only the $Z_b^+$ component. 

The comparison between the $Z_b ^+$ production and absorption cross sections, 
shown in Figs.~\ref{CrSecZb} and \ref{CrSecZbInv}, suggests that the 
$Z_b$-production cross sections are smaller than the absorption ones by a  
factor about 3-100, depending on the specific channel and region of energy. 
This difference between production and absorption cross sections can be 
accounted for by  kinematic effects.  



In what follows we use the results reported above to
compute the thermally averaged cross sections for the
$Z_b$ production and absorption reactions.




\section{Cross Sections averaged over the thermal distribution  }

\label{ThermalAvCrossSection}




Let us introduce the cross section averaged over the thermal distribution 
for a reaction involving an initial two-particle state going into two final 
particles $ab \to cd$. It is given by~\cite{ChoLee1,XProd2,Koch}
\begin{eqnarray}
 \langle \sigma_{ab \to cd} \, v_{ab} \rangle  
 & = &  \frac{\int d^3\vec{p}_a d^3\vec{p}_b f_a(\vec{p}_a) 
f_b(\vec{p}_b) \sigma_{ab \to cd}v_{ab} }{\int d^3\vec{p}_a 
d^3\vec{p}_b f_a(\vec{p}_a) f_b(\vec{p}_b)} 
\label{ThermalCrossSection}
\end{eqnarray}
where $f_a$ and $f_b$ are Bose-Einstein distributions, $\sigma_{ab \to cd}$ 
are the cross sections evaluated in Section~\ref{CrossSection}, $v_{ab}$ 
represents the relative velocity of the two interacting particles  
$a$ and $b$.

\begin{figure}[!ht]
    \centering
        \includegraphics[width=8.0cm]{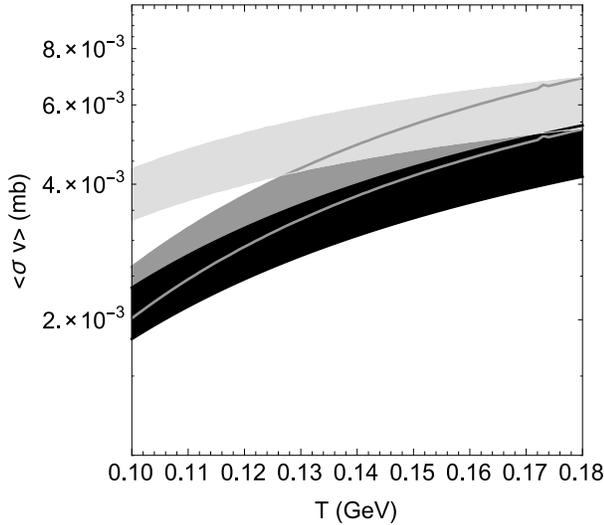}
        \caption{Thermally averaged cross section to the processes $ \bar{B} B \rightarrow \pi Z_b $ (medium shaded region), $ \bar{B}^* B \rightarrow \pi Z_b $ (dark shaded region) and $ \bar{B}^* B^* \rightarrow \pi Z_b $ (light shaded region), as a function of temperature.}
\label{AvCrSecZb}
\end{figure}

\begin{figure}[!ht]
    \centering
         \includegraphics[width=8.0cm]{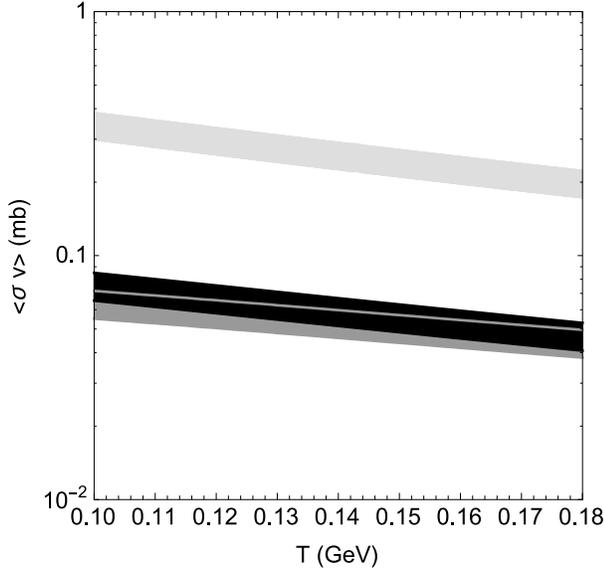}
        \caption{Thermally averaged cross section to the processes $    \pi Z_b \rightarrow \bar{B}  B $ (medium shaded region), $  \pi Z_b \rightarrow \bar{B}^* B $ (dark shaded region) and  $  \pi Z_b \rightarrow \bar{B}^*  B^*$ (light shaded region), as a function of temperature.}
\label{AvCrSecZbInv}
\end{figure}

In Figs.~\ref{AvCrSecZb} and ~\ref{AvCrSecZbInv} we plot the thermally 
averaged cross sections for  $Z_b$ absorption and production 
respectively, via the processes discussed in the previous Section. All 
the three  production reactions considered have similar  
magnitudes, especially at high temperatures. 
The absorption processes with $B \bar{B}$ and $B^* \bar{B}$ final  states  
have comparable cross sections and these are smaller than in the case with 
$B^* \bar{B}^* $ final state (by a factor about 2-3). 
Comparing Figs.~\ref{AvCrSecZb} and ~\ref{AvCrSecZbInv} we observe that absorption is stronger than production by a factor ranging from $20$ to $100$  in the energy region of interest.


In the next section we use these thermally averaged cross sections as input in the rate equation and study  the time evolution of the $Z_b$  abundance.



\section{Time evolution of $Z_b$ abundance }

\label{abundance}




We complete the present investigation with the study  of the time evolution  of the $Z_b$ abundance in hadronic matter, using the thermally averaged cross sections estimated in the previous  section. More precisely,
we investigate the influence of $\pi - Z_b $  interactions on the abundance of $Z_b$ during the hadronic stage of heavy ion collisions.     
The momentum-integrated evolution equation for the 
abundances of particles included in processes previously discussed
~\cite{ChoLee1,XProd2,Koch} reads 
\begin{eqnarray} 
\frac{ d N_{Z_b} (\tau)}{d \tau} & = & \sum_{b,b^{\prime}} 
\left[ \langle \sigma_{b b^{\prime} \rightarrow \pi Z_b } 
v_{b b^{\prime}} \rangle n_{b} (\tau) N_{b^{\prime}}(\tau)
 \right. \nonumber \\ & &  \left. 
- \langle \sigma_{ \pi Z_b \rightarrow b b^{\prime} } v_{ \pi Z_b} 
\rangle n_{\pi} (\tau) N_{Z_b}(\tau)  \right], 
\label{rateeq}
\end{eqnarray}
where $N_{Z_b} (\tau)$, $N_{b^{\prime}}(\tau)$,  $ n_{b} (\tau)$ and 
$n_{\pi} (\tau)$ are the abundances of $Z_b$, of  bottomed 
mesons of type $b^{\prime}$, of  bottomed mesons of type $b$ and of  
pions  at proper time $\tau$, respectively. We can see in the above equation that the $Z_b$ abundance at a proper time $\tau$ 
depends on the $\pi Z_b$ dissociation rate through the processes 
discussed previously, and also on the  $\pi Z_b$  production rate 
from  the inverse processes.

To solve Eq.~(\ref{rateeq}) we assume that  the pions and bottomed 
mesons in the reactions contributing to the abundance of $Z_b$ are   
in equilibrium. Accordingly, $ n_{b} (\tau)$, $N_{b^{\prime}}(\tau)$ 
and $n_{\pi} (\tau)$ can be written as~\cite{ChoLee1,XProd2,Koch}
\ben
n_{b} (\tau) &  \approx & \frac{1}{2 \pi^2} \, \gamma_{b} \,  
g_{B^{(\ast)}} \,  m_{B^{(\ast)}}^2 \,  T(\tau) \, 
K_{2}\left(\frac{m_{B^{(\ast)}} }{T(\tau)}\right), \nonumber \\
N_{b^{\prime}}(\tau) & \approx & \frac{1}{2 \pi^2} \, \gamma_{b} \,  
g_{B^{(\ast)}} \,  m_{B^{(\ast)}}^2  \, T(\tau) \,  V(\tau) \, 
K_{2}\left(\frac{m_{D^{(\ast)}} }{T(\tau)}\right) ,\nonumber \\
n_{\pi} (\tau) &  \approx & \frac{1}{2 \pi^2} \, \gamma_{\pi} \,  
g_{\pi} \, m_{\pi}^2 \,  T(\tau) \, 
K_{2}\left(\frac{m_{\pi} }{T(\tau)}\right), 
\label{densities}
\een
where $\gamma _i$ and $g_i$ are the fugacity factor and the  degeneracy of  particle $i$ respectively. As it can be seen in Eq.~(\ref{densities}), 
the time dependence in Eq.~(\ref{rateeq}) enters through the 
parametrization of the temperature $T(\tau)$ and volume $V(\tau)$ profiles suitable to 
describe the dynamics of the hot hadron gas after the end of the quark-gluon plasma phase.  
As in Refs.~\cite{ChoLee1,XProd2,Koch},  we assume the $\tau$ dependence of $V(\tau)$ and $T$ to be given by
\ben
V(\tau) & = & \pi \left[ R_C + v_C \left(\tau - \tau_C\right) + 
\frac{a_C}{2} \left(\tau - \tau_C\right)^2 \right]^2 \tau c , \nonumber \\
T(\tau) & = & T_C - \left( T_H - T_F \right) \left( \frac{\tau - 
\tau _H }{\tau _F - \tau _H}\right)^{\frac{4}{5}} .
\label{TempVol}
\een
These expressions are based on the boost invariant Bjorken picture 
with an accelerated transverse expansion. 
Since we focus on central Au-Au collisions at 
$\sqrt{s_{NN}} = 200$ GeV, in the above equation 
$R_C = 8.0$ fm denotes the final  size of the quark-gluon plasma, 
while $v_C = 0.4 \, c$  and  $a_C = 0.02 \,  c^2$/fm are its transverse 
flow velocity and transverse acceleration at $\tau_C = 5.0$  fm/c~
\cite{ChoLee1,XProd2,ChenPRC}.  The critical temperature of the  
quark gluon plasma to hadronic matter transition  is $T_C=175$ MeV; 
$T_H = T_C = 175$ MeV is the temperature of the hadronic matter at 
the end of the mixed phase,  at $\tau_H = 7.5$ fm/c. The freeze-out 
takes place at the freeze-out time $\tau_F = 17.3$ fm/c, 
when the temperature drops to  $T_F = 125$ MeV.

We assume that the total number of bottom quarks in bottomed hadrons 
is conserved during the  production and dissociation reactions, and 
that the total number of bottom quark pairs  produced at the initial 
stage of the collisions at RHIC is 0.02, yielding the bottom 
quark fugacity factor $\gamma _b \approx 2.2 \times 10^6 $ in 
Eq.~\eqref{densities} \cite{exhic1,exhic2}. In the case of pions,  
their total number at freeze-out is assumed to be 926
~\cite{ChoLee1,XProd2,ChenPRC}.

In the present work we study the yields obtained for the $Z_b $   
abundance within two different approaches: the statistical and the 
coalescence models. In the statistical model, hadrons are produced in 
thermal and chemical equilibrium, according to the expression of the 
abundance in Eq.~(\ref{densities}) corresponding to the hadron considered. 
Therefore,  the $Z_b $ yield at the end of the mixed phase (produced 
from quark-gluon plasma) is 
\ben
 N_{Z_b(Stat)} ^{0 } =N_{Z_b(Stat)} (\tau_H)\approx  2.1 \times 10^{-8}.
\label{NZstat}
\een
Notice, however, that this model does not contain any information 
related to the internal structure of the $Z_b $.        
In the coalescence model the 
determination of the yield of a certain hadron is based on the overlap 
of the density matrix of the constituents in an emission source with 
the Wigner function of the produced particle.  This model contains 
information on the internal structure of the considered hadron, such as 
angular momentum, multiplicity of quarks, etc.                             
Then, following Refs.~\cite{exhic1,exhic2,ChoLee1,XProd2}, the number 
of $Z_b$'s produced  at the end of the mixed phase can  be written as :
\ben
N_{Z_{b}} ^{Coal} & \approx & g_{Z_{b}} \prod _{j=1} ^{n} \frac{N_j}{g_j} 
\prod  _{i=1} ^{n-1} 
\frac{(4 \pi \sigma_i ^2)^{\frac{3}{2}} }{V (1 + 2 \mu _i T \sigma _i ^2 )} 
\nonumber \\
& & \times
\left[ \frac{4 \mu_i T \sigma_i ^2 }{3 (1 + 2 \mu _i T \sigma _i ^2 ) }
\right]^{l_i}, 
\label{NZCoal}
\een
where $g_j$ and $N_j$ are the degeneracy and number of the $j$-th constituent of the $Z_b$  and 
$\sigma _i = (\mu _i \omega)^{-1/2}$. The quantity $\omega $ is the 
oscillator frequency (assuming an harmonic oscillator Ansatz for the hadron internal structure) and $\mu$ the reduced mass, given by  
$\mu ^{-1} = m_{i+1} ^{-1}+ \left(\sum_{j=1} ^{i} m_j \right)^{-1}$. 
Finally,  the angular momentum of the system, $l_i$, is 0 for an $S$-wave,  
and 1 for a $P$-wave. In order to calculate the  $Z_b$ yield within 
the coalescence model, we consider it as a tetraquark state.  
Therefore, according to  the coalescence model for $Z_b $ as a tetraquark, the time evolution of the abundance ($  N_{Z_{b}} $) is determined by solving Eq.~(\ref{rateeq}), with initial condition, at $\tau_H = 7.5$ fm/c,  given by
\ben
 N_{Z_{b}(4q)}  ^{0 } =N_{Z_{b}(4q)} (\tau_H)
  \approx  2.1 \times 10^{-9}.  \label{NZ4q}
\een
The comparison between the values of $ N_{Z_{b}(Stat)}  ^{0 } $  
and $ N_{Z_{b}(4q)}  ^{0}$ in Eqs.~(\ref{NZstat}) and~(\ref{NZ4q}) indicates that the number of $Z_b $'s (produced at the end of the mixed phase) calculated with the statistical model is greater than the  four-quark state (formed by quark coalescence)  by one order of magnitude. 

\begin{figure}[!ht]
	\centering
	\includegraphics[width=8.0cm]{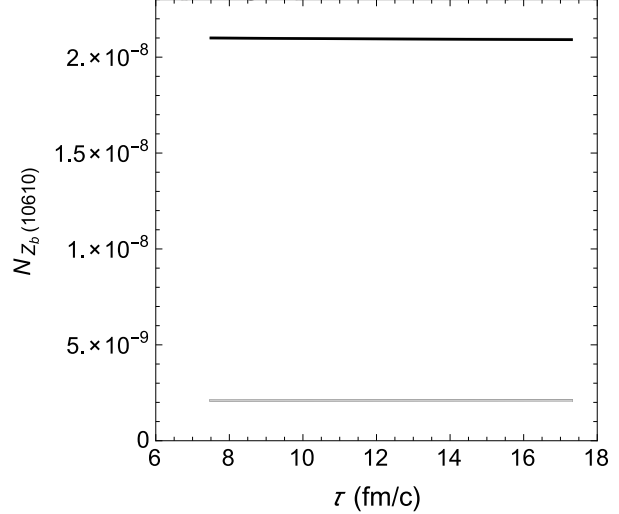}
	\caption{Time evolution of the $Z_b$ abundance as a function of 
the proper time in central Au-Au collisions at $\sqrt{s_{NN}} = 200$ GeV.   
Dark and light shaded bands represent the evolution with the number of  
$Z_b$'s produced  at the end of the mixed phase calculated using  
statistical and four-quark coalescence models, respectively.}
	\label{TimeEvolZb}
\end{figure}

In Fig.~\ref{TimeEvolZb} we show the time evolution of the $Z_b$ abundance as a function of the proper time in central Au-Au collisions at 
$\sqrt{s_{NN}} = 200$ GeV, using $ N_{Z_{b}(Stat)}  ^{0 } $ and 
$ N_{Z_{b}(4q)}  ^{0}$ as initial conditions.  
The results suggest that the  interactions between the $Z_b$'s and the pions during the hadronic stage of heavy ion collisions do not produce any relevant change in the  $Z_b$ abundance, i.e. there is an approximate equilibrium between production  and absorption and  the number of  $Z_b$'s  throughout the hadron gas phase remains nearly constant.



\section{Conclusions}

\label{Conclusions}



In this work we have studied the $Z_b (10610)$ abundance in a hot 
pion  gas produced in heavy ion collisions. Effective Lagrangians have 
been used to calculate the thermally  averaged  cross sections of the 
processes $B^{(*)} + \bar{B}^{(*)} \to \pi + Z_b (10610)$,  as well as 
of the corresponding inverse processes. We have found that the magnitude of the thermally averaged cross sections for  
the dissociation and for the production reactions differ by 
factors up to $100$, depending on whether the considered channel includes or not the $B^{(*)} + \bar{B}^{(*)}$ state.

With the thermally averaged cross sections we have solved the rate equation  to determine the time evolution of the  $Z_b (10610)$ multiplicity. The results suggest that the $Z_b $ yield  is not significantly 
affected by the interactions with the pions and hence the number of 
$Z_b$'s remains essentially unchanged during the hadron gas phase.

\begin{acknowledgements}


The authors would like to thank the Brazilian funding agencies CNPq and 
FAPESP (contract number 12/50984-4 and 17/07278-5) for financial support.

\end{acknowledgements}


\end{document}